\documentclass[usenatbib]{mn2e}
\usepackage{graphicx}
\usepackage[normalem]{ulem}
\newcommand{\mnras}{MNRAS}

\def\apj{ApJ}
\def\apjl{ApJL}
\def\apjs{ApJS}
\def\aj{AJ}

\def\mnras{MNRAS}
\def\nat{Nature}

\def\prl{PRL}
\title[Morphological Properties of Superclusters of Galaxies]{Morphological Properties of Superclusters of Galaxies}
\author[Costa-Duarte, M.V., Sodr\'e, L., Durret, F.]{Costa-Duarte, M.V.$^{1}$\thanks{e-mail:mvcduarte@astro.iag.usp.br}, Sodr\'e Jr., L.$^{1}$,Durret, F.$^{2,3}$\thanks{This file has been amended to
highlight the proper use of \LaTeXe\ code with the class file.}\\
$^{1}$Instituto de Astronomia, Geof\'isica e Ci\^encias Atmosf\'ericas, USP, R. do Mat\~ao 1226, 05508-090, S\~ao Paulo, Brazil\\
$^{2}$UPMC Universit\'e Paris 06, UMR~7095, Institut d'Astrophysique de
Paris, F-75014, Paris, France \\
$^{3}$CNRS, UMR~7095, Institut d'Astrophysique de Paris, F-75014, Paris,
France}

\begin{document}

\pagerange{\pageref{firstpage}--\pageref{lastpage}} \pubyear{2010}

\maketitle

\label{firstpage}

\begin{abstract}
We studied superclusters of galaxies in a volume-limited sample extracted from the Sloan Digital Sky Survey Data Release 7 (SDSS/DR7) and from mock catalogues based on a semi-analytical model of galaxy evolution in the Millenium Simulation. A density field method was applied to a sample of galaxies brighter than $M_r= -21+5 \log h_{100}$ to identify superclusters, taking into account selection and boundary effects. In order to evaluate the influence of threshold density, we have chosen two thresholds: the first maximizes the number of objects (D1), and the second constrains the maximum supercluster size to $\sim$120~h$^{-1}$Mpc (D2). We have performed a morphological analysis, using Minkowski Functionals, based on a parameter which increases monotonically from filaments to pancakes. An anti-correlation was found between supercluster richness (and total luminosity or size) and the morphological parameter, indicating that filamentary structures tend to be richer, larger and more luminous than pancakes in both observed and mock catalogues. We have also used the mock samples to compare supercluster morphologies identified in position and velocity spaces, concluding that our morphological classification is not biased by the peculiar velocities. Monte Carlo simulations designed to investigate the reliability of our results with respect to random fluctuations show that these results are robust. Our analysis indicates that filaments and pancakes present different luminosity and size distributions. 

\end{abstract}


\section{Introduction}

It is well known that galaxies are not randomly distributed in the Universe, with high-density regions being observed as large-scale structures and low-density regions as voids. Under the current $\Lambda$CDM cosmological paradigm, the evolution of these structures started in the early Universe from primordial density fluctuations just after
inflation, leading to the observed cosmic web. At very large scales, of  tens of Mpc, clusters, groups, and even pairs or isolated galaxies are disposed in very large associations, sometimes of filamentary or planar
structure. These associations are the largest non-virialized structures in the Universe: superclusters of galaxies. Their dynamical future is still uncertain, but in a dark energy dominated Universe most of them may evolve to \textit{island universes}, single, isolated and highly concentrated mass clumps \citep{Araya-Meloetal2009}.

The study of very large scale structures started with \cite{deVaucouleurs1953}, who identified a high-density region in the galaxy distribution on the sky, nowadays known as the Local Supercluster. \cite{Abell1958} also helped to unveil large-scales through his catalogue of clusters of galaxies identified in the Palomar Observatory photographic plates; he defined superclusters as clusters of clusters of galaxies. Larger surveys were carried out and, as a consequence, the distribution of galaxies in the local Universe could be studied in detail. The Harvard Center for Astrophysics (CfA) survey measured the redshift of a sample of galaxies brighter than 14.5 \citep{Huchraetal1983}, showing the filamentary distribution of galaxies, with galaxy clusters at the connection of the filaments.  These redshifts allowed to constrain the cosmological model. Indeed, \cite{Efstathiouetal1990} showed, from the analysis of large-scale galaxy clustering in the IRAS survey, that a cosmological constant was required to explain the galaxy distribution in the framework of the CDM model. 

The data derived from redshift surveys later allowed to study the properties of individual superclusters, like Pisces-Cetus \citep{Tully1988} and Shapley \citep{Proustetal2006}, as well as those of the whole population of superclusters, revealing that they tend to have an  elongated morphology and extensions up to $\sim$100~h$^{-1}$~Mpc \citep{BahcallSoneira1984}. Recent works reveal extensions
up to   110-130h$^{-1}$Mpc \citep{Pandeyetal2010}. 
The study of superclusters highly benefited from the \textit{2 degree Field Galaxy Redshift Survey} (2dFGRS) \citep[][]{Colless2001} and the \textit{Sloan Digital Sky Survey} (SDSS) \citep[][]{Abazajianetal2009}. 

Indeed, more complete studies of large-scale structures were possible using these large redshift surveys. Jaan Einasto's group used the 2dFGRS data to generate a catalogue of superclusters \citep[][hereafter E07a]{Einastoetal2007a}\defcitealias{Einastoetal2007a}{E07a}. They also compared observed superclusters to simulated ones \citep[][hereafter E07b]{Einastoetal2007b} and studied the spectral properties of galaxies within superclusters  \citep{Einastoetal2007c}. Their main results indicate that the overall properties of simulated and observed superclusters present good agreement with each other, but their luminosity and multiplicity (number of galaxies) distributions seem to be different. Additionally, they found that galaxy morphology in superclusters depends on their richness, with rich superclusters presenting an early-type fraction slightly higher than poor superclusters. In another series of papers (Einasto et al. 2007d,e), these authors have studied the richest superclusters identified in the observations. Comparing the clumpiness of simulated and observed superclusters, they conclude that the clumpiness of galaxies in simulations is different from that observed \citep{Einasto2007-Morph}, and that the global and local environments are quite important for galaxy morphology and star formation activity \citep{Einasto2007-GalPop}.

Since superclusters are non-virialized structures, they present a variety of morphologies \citep{West1989,Plionisetal1992}. Several studies have used shape statistics \citep{Sahnietal1998} and Minkowski Functionals \citep{Meckeetal1994} to determine topological and geometrical properties useful for morphological analysis. The SDSS and PSCz (Saunders et al. 2000) superclusters seem to have a prevalence of filamentary structures \citep{Basilakos2003}, as well as a concordance with the $\Lambda$CDM model of large-scale structure formation \citep{Basilakosetal2001}(hereafter B01). A comparison of observed and simulated superclusters showed that simulated superclusters are very similar to those observed, but the number-density of very luminous superclusters seems to be higher in observations than in simulations \citep[][hereafter E06]{Einastoetal2006}. 

Morphological studies suggest that galaxies are found in two distinct classes 
of structures at very large scales: filaments and pancakes. Using the 
shapefinder technique \citep{Sahnietal1998}, B01 applied this approach 
to distinguish between these two classes using the so called shape-spectrum. 
Further works used filament features to constrain the galaxy clustering, 
since the bias parameter is also sensitive to filamentarity 
\citep{Bharadwaj-Pandey2004}. Using galaxy luminosities and colours, 
\cite{Pandey-Bharadwaj2006} found a dependence between galaxy properties and 
filamentarity, proposing a scenario where elliptical galaxies are 
predominantly in dense regions, while spiral galaxies are distributed along filaments. A strong spatial alignment between clusters and host superclusters in
large filaments was found in N-body simulations 
\citep{Basilakosetal2006,Lee-Edvard2007}. 

Here we present a study of supercluster morphologies through the study of the galaxy distribution in volume-limited samples extracted from SDSS Data Release~7 \citep{Abazajianetal2009}. We use a kernel-based density field method to identify the superclusters and Minkowski Functionals to quantify their shape.

This paper is organized as follows. Section~\ref{data} presents the data used here as well as our method to deal with selection effects. In section~\ref{densityfield} we describe the kernel-based density field method used to identify superclusters as well as the criteria to classify enhanced regions as superclusters taking into account the selection and boundary effects. In section~\ref{morphclass} the morphological classification is described and  in section~\ref{resultsanddiscussion} we discuss the morphology of observed and simulated superclusters.  Finally, in section~\ref{conclusion} we summarize the main conclusions of this paper. In Appendix \ref{apend_a} we discuss the sensitivity of our 
supercluster identification and morphological analysis to the adopted kernel. 

When necessary, distances were calculated assuming the following cosmology: $\Omega_m=0.3$, $\Omega_\Lambda=0.7$ and Hubble parameter $H_0$=100~h$_{100}^{-1}$~km~s$^{-1}$Mpc$^{-1}$.

\section[]{Data}
\label{data}
The analysis presented in this paper is based on a volume-limited galaxy sample  extracted from the Sloan Digital Sky Survey Data Release Seven (SDSS-DR7) \citep{Abazajianetal2009}. We have considered galaxies with measured radial velocities and with absolute magnitudes in the $r$ band brighter than $-21 + 5 \log h$ in the redshift range $0.04 \le z \le 0.155$. Absolute magnitudes were calculated with k-corrections obtained with the code $KCORRECT$ $v4.1.4$ and with a specific SDSS package provided by \cite{Blanton&Roweis2007}. Since superclusters may extend over several degrees on the sky, we have considered only galaxies within stripes 10 to 37 to assure  a large continuous area on the sky. The total number of galaxies selected is 120,013.

In Section 5 we shall compare some of our results with numerical simulations. For this we have used simulated light-cones produced by \cite{Crotonetal2006}, based on a semi-analytic galaxy evolution model applied to the output of the Millenium Simulation \citep{Springeletal2005}. We have selected the four light-cones with parameters suited for SDSS: SDSS\_SAcone\_012\_000, SDSS\_SAcone\_012\_100, SDSS\_SAcone\_120\_000, and SDSS\_SAcone\_201\_000. Each covers an area of 60 $\times$ 30 deg$^2$ and the simulated galaxies were selected following the same criteria adopted in the selection of our volume-limited sample of SDSS galaxies. The number of simulated galaxies selected in the four light-cones is 99,850.

\section{The Density Field Method}
\label{densityfield}

Superclusters are sometimes defined as large-scale overdensity regions in the galaxy distribution \citep{deVaucouleurs1953}. Adopting this definition, the density field method represents a convenient way to identify these structures (e.g., E07a, B01). In this section we describe how we define the density field of a sample of galaxies. 

\subsection{The density field}
\label{kernel}

Firstly, using the equatorial coordinates $(\alpha,\delta)$ and the redshift $z$ of each galaxy, we calculated its cartesian coordinates as
\begin{eqnarray}
x&=&d_c \cos(\delta)\cos(\alpha) \nonumber \\
y&=&d_c \cos(\delta)\sin(\alpha) \nonumber \\
z&=&d_c \sin(\delta) 
\end{eqnarray}
where $d_c(z)$ is comoving distance of the galaxy.

The luminosity density of the galaxy distribution, $D(\mathbf{r})$, is calculated through the kernel approach. At a certain point ${\mathbf r}$ in space it is given by
\begin{center}
\begin{equation}
D(\mathbf{r})=\sum_i K(|\mathbf{r-r_i}|,\sigma)L_i W_{i}(\mathbf{r_i}),
\end{equation}
\end{center}
where $K(r,\sigma)$ is the kernel used to smooth the galaxy distribution, $L_i$ is the luminosity of the $i$-th galaxy (at position ${\mathbf r_i}$) and $W_i({\mathbf r_i})$ is a weight which takes into account selection effects (discussed in the next section). 

We adopt here Epanechnikov's kernel, which minimizes the asymptotic mean
integrated squared error \citep[e.g.][]{silvermann1986} and is defined as
\begin{eqnarray}
 \ \ \ \ \ \ \ \ \ \ \ \ \ \ \ \ K(r,\sigma)=\left\{ 
\begin{array}{c}
\frac{3}{4}[1-(r/\sigma)^2]\ \ \ \ r\leq\sigma\\ 
0\ \ \ \ \ \ \ \ \ \ \ \ \ \ \ \ \ \ \ \ r>\sigma\\
\end{array}
\right.
\label{kernel2}
\end{eqnarray}
 Here $\sigma$ is the smoothing parameter. We have adopted $\sigma$=8~h$^{-1}$Mpc. The reason is that the number density of galaxies in our sample (see next section) is $\bar{n}=2.1\times10^{-3}$(h$^{-1}$Mpc)$^{-3}$, corresponding to a mean distance between galaxies of $\sim$8~h$^{-1}$Mpc. As shown later, this choice leads to a density field relatively insensitive to peculiar velocities. The density field is sampled in a 3D grid with cells of side $l_{cell}$=4~h$^{-1}$Mpc. 

To identify structures in the density field, it is necessary to define a density threshold to separate high-density regions (e.g., superclusters) from low-density regions (e.g., voids). In this way, we have rejected all grid points below the threshold. Afterwards, a friends-of-friends algorithm was used to connect nearby high-density grid points, assigning them to single objects. The linking-length used is equal to the diagonal of the cell grid, i.e., $l_{fof}=\sqrt{3}~l_{cell}\simeq$7~h$^{-1}$Mpc. Only objects with more than 10 galaxies and volume larger than two grid cells, $V_{min}$=$2(l_{cell})^3$=128(h$^{-1}$Mpc)$^3$, will be considered in the analysis \citepalias[][]{Einastoetal2007a}.  

Note that about 0.3\% of the DR7 imaging footprint area are marked as holes. 
In these regions we have used bilinear 
interpolation to obtain the density field, considering only grid points at
a given redshift more distant that 8 h$^{-1}$Mpc from the hole borders.
Since the area occupied by the holes is small, and most of the holes are in 
low-density regions, it can be verified that this
procedure has a negligible impact on our results. 

\begin{figure}
\includegraphics[scale=0.5]{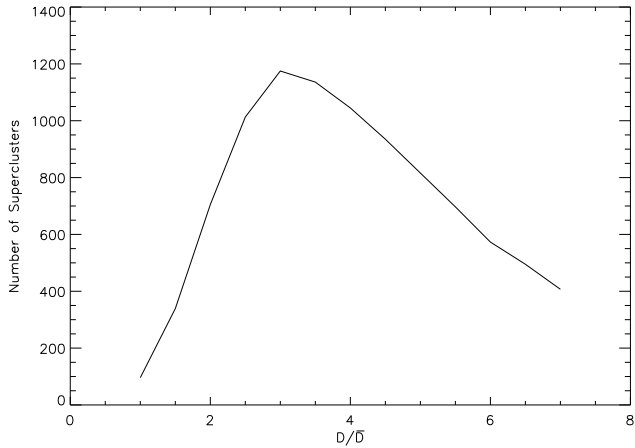} 
\caption{Number of superclusters as a function of the threshold density (in units of the mean density).}
\label{nscxdens}
\end{figure}

There is no natural value for the threshold density. In Figure~\ref{nscxdens} we present the number of structures (hereafter called superclusters) as a
function of the threshold in units of mean density ($\bar{D}$), computed with the selection function discussed in {\bf 3.2}. For low threshold density values, percolation links distinct structures and consequently the number of superclusters is low. At high threshold values, only high density objects are identified, also resulting in a low number of structures. In this work we have adopted two distinct values for the threshold density. The first one, $D_{thresh} = 3 \times \bar{D}$ (hereafter D1), is the value which maximizes the number of superclusters.  The second one, $D_{thresh} = 6 \times \bar{D}$ (hereafter D2), was chosen such as the largest superclusters present an extension of $\sim$120~h$^{-1}$Mpc, as adopted by \citetalias{Einastoetal2007a}. This length consists of the diagonal of the box which contains all galaxies of the supercluster, i.e., $l=\sqrt{\Delta x^2+\Delta y^2+\Delta z^2}$.

Two important features of superclusters can be defined here: their richness and total luminosity. The richness can be written as
\begin{equation}
 R=\sum^{N_{gal}}_i W_i 
\end{equation}
where $N_{gal}$ represents the number of galaxies of the supercluster and $W_i$ is the selection effect correction of the $i$-th galaxy. The total supercluster luminosity- actually the expected luminosity above the magnitude limit- is defined as 
\begin{equation}
 L_{tot}=\sum^{N_{gal}}_i L_i W_i
\end{equation}
where $L_i$ represents the luminosity of the $i$-th galaxy.

It is worth mentioning that our results are not too sensitive to the choice
of the smoothing kernel. We present in Appendix \ref{apend_a} a summary of results
obtained with a truncated Gaussian kernel, showing that our estimates of 
supercluster parameters are indeed very robust.

\subsection{A model for the selection function}
The selection function aims at correcting for galaxies brighter than our 
magnitude limit that, for a reason or another, were not included in the sample.

\begin{figure}
\centering
\includegraphics[scale=0.5]{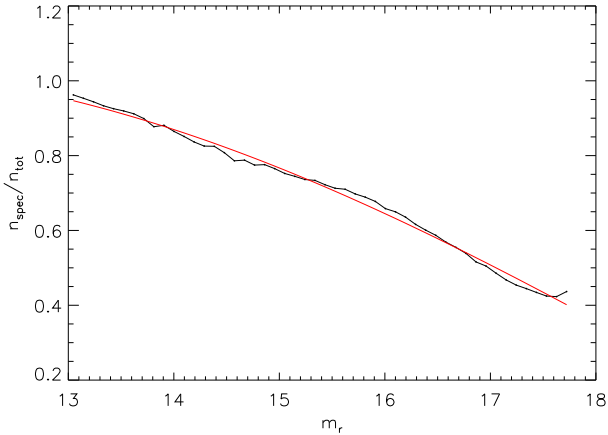} 
\caption{Fraction of galaxies with observed spectra as a function of apparent r-band magnitude in the range $13.0<m_r<17.77$. The red line represents the fourth order polynomial fitted to the observed trend.}
\label{completeness}
\end{figure}

Indeed, our magnitude limited sample is affected by incompleteness due to
fiber collisions in the spectroscopic survey. Consequently, although the 
nominal magnitude limit of the SDSS Main Galaxy Sample MGS) is $m_r = 17.77$, 
not all galaxies brighter than this limit were observed. 
There is a minimum distance between fiber allocations by the SDSS 
spectrographs of about 55 arcsec \citep{Straussetal2002} and some galaxies 
within the MGS photometric limits do not have spectroscopy because they are 
closer than 55 arcsec to a galaxy to which a fiber was allocated.  
This spectroscopic incompleteness depends on the apparent magnitude 
since fainter galaxies are more affected. This is shown in 
Figure~\ref{completeness}, which presents  
the fraction of galaxies with spectroscopy as a function of the apparent 
magnitude $m_r$. 

The spectroscopic incompleteness leads to a radial selection effect, since 
galaxies with higher apparent magnitude tend to be at higher redshifts.
Since in this case the shot-noise increases with distance, the coupling 
between this radial  effect and a constant smoothing parameter 
introduces an additional distortion in the density field 
\citep{GaztanagaYokoyama1993,Sejlaketal2009}. This bias leads to an overestimation of the
density with increasing redshift (B01). 

To deal with these effects, we have adopted a simple model for the selection 
function, with two components. The first one depends on the apparent magnitude 
($S_1(m_r)$) and the second one on the redshift, or comoving distance 
($S_2(d_c)$).  The selection function is thus defined as
\begin{equation}
 S(m_r,d_c)=S_1(m_r)S_2(d_c)
\end{equation}
and is related to the weight $W$ as 
\[ W = S(m_r,d_c)^{-1}  \]

The  apparent magnitude component can be defined as $S_1(m_r) = n_{spec}/n_{tot}$
and we model the trend seen in Figure~\ref{completeness} with a fourth order 
polynomial:
\begin{eqnarray}
S_1(m_r)=0.588605-1.941834m_r+0.419142m_r^2 \nonumber \\
-0.029956m_r^3+0.000724m_r^4
\end{eqnarray}

\begin{figure}
\centering
\includegraphics[scale=0.5]{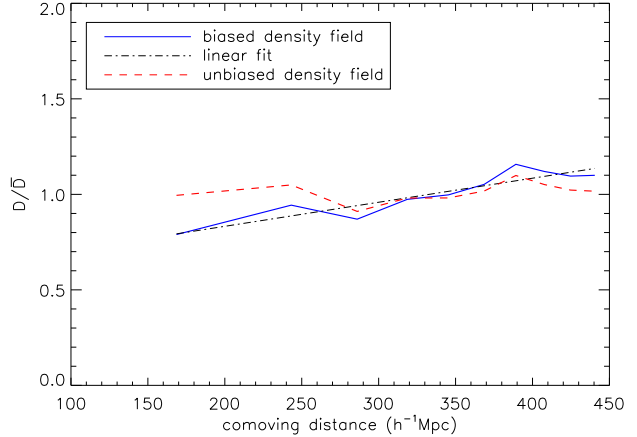} 
\caption{The blue continuous line shows the mean density in equal
  volume regions as a function of the comoving distance when only the
  magnitude selection effect is taken in to account. The slight
  dependence of the density with distance shows evidence for a radial
  bias. The blue dot-slash-dot line is the linear fit adopted to model
  this radial bias and the red dashed line represents the mean density
  after correcting the density field by the magnitude and radial
  selection effects.}
\label{shot-noise}
\end{figure}

To estimate the radial component of the selection function we, initially,
calculated the mean of the density field grid points in ten regions 
with same volume, taking into account only the apparent magnitude 
incompleteness (i.e., assuming $S_2(d_c)=1$). As shown in Figure 
\ref{shot-noise}, the mean density of each region increases with
redshift, reflecting the bias mentioned above. However, a comparison with Figure 3 of B01 indicates that the effect here, is significantly less severe in our volume limited sample than in magnitude limited samples.
To correct for this effect, we model the dependence of the mean density
with comoving distance as $S_2(d_c)=a*d_c+b$ with $a=0.0025$ and $b=0.1565$ for distances in Mpc.  Taking $S_2(d_c)$ into account, the radial trend in
the mean value of the density field disappears, as shown in Figure 
\ref{shot-noise}.

\subsection{Boundary Effects}
Due to the large sizes of superclusters and to the limited volume of our galaxy sample, care should be taken to avoid boundary effects that can affect the analysis described in the next sections.

\begin{figure}
\begin{center}
\includegraphics[scale=0.5]{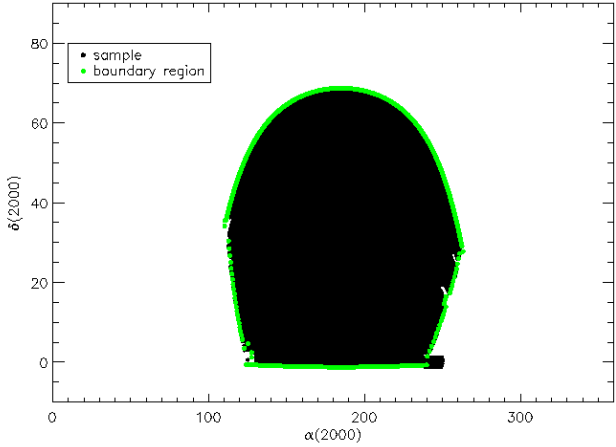} 
\caption{Region occupied by our sample on the sky (in black). Green dots represent the boundary of the region.} 

\label{radec-boundary}
\end{center}
\end{figure}

With this aim, we considered only structures where all galaxies have comoving transversal distances from any volume boundary border larger than $\sigma=$ 8~h$^{-1}$Mpc. Figure~\ref{radec-boundary} shows the region occupied by our initial sample (in black) as well as its boundary points (in green). Superclusters in the region with $240^{\circ}<\alpha<253^{\circ}$ and $-2^{\circ}<\delta<+2^{\circ}$ were excluded from our sample to avoid boundary effects. 
Excluding superclusters at the boundary, our final supercluster sample has 880 structures above the threshold D1 and 409 structures above D2. Their main properties are shown in Table~1.

\section{Morphological Description of Superclusters}
\label{morphclass}
Since superclusters of galaxies are not virialized structures, they present a large variety of shapes. In this section we present the shapefinder method which we adopted to describe the morphology of these structures. First we discuss the ellipsoidal model to describe the structures and then we use Minkowski Functionals to obtain a morphological parameter. 

\subsection{The ellipsoidal model}
\label{fit-ellipsoid}
A simple description of the tri-dimensional morphology of a body can be 
obtained through its best-fit tri-axial ellipsoid. This model has 8 free parameters: three for the centroid of the structure, two for its orientation and three for the semi-axes $a_1$, $a_2$ and $a_3$.

The parameters of the ellipsoid can be inferred from the inertia tensor, i.e., the matrix of second-order moments of particle positions,
\begin{equation}
\label{inertiamomentum}
I_{ij}=\sum_k L_k W_{k} x_i^k x_j^k,
\end{equation} 
where $x_i^k$ represents de $i$-th coordinate of the $k^{th}$ galaxy with respect to the object centroid. The matrix $I_{ij}$ can be diagonalized and the diagonal elements are proportional to the best-fit ellipsoid semi-axes \citep[e.g.,][]{Jang-Condell&Hernquist2001,Plionisetal1991,Kolokotronisetal2002},
\begin{eqnarray}
\ \ \ \ \ \ \ \ \ \ \ \ \ \ \ \ \ \ \ I_1&=&\frac{\sum_i L_i W_{i}}{5}(a_2^2+a_3^2) \nonumber \\
I_2&=&\frac{\sum_i L_i W_{i}}{5}(a_1^2+a_3^2) \nonumber \\
I_3&=&\frac{\sum_i L_i W_{i}}{5}(a_1^2+a_2^2).
\end{eqnarray}
Solving the system of equations above, the three semi-axes are determined, with the assumption that $a_1 \ge a_2 \ge a_3$. 

\subsection{Minkowski Functionals}

Minkowski Functionals (MFs) represent an important tool to describe structures
and objects, since they characterize their geometry. We follow here the formalism of \cite{Sahnietal1998}, which uses ellipsoidal models for the morphological description of the objects. 

Having as input parameters the semi-axes $a_1$, $a_2$, $a_3$ obtained in the previous section, we can determine, for an object or isodensity contour, four parameters: the volume (V), the surface (S), the
integrated mean curvature (C) and the integrated Gaussian curvature
(${\cal G}$), also called \textit{genus}.

The parametric equation for an ellipsoid with semi-axes $a_1$, $a_2$ and $a_3$ 
can be written as
\begin{center}
\begin{equation}
{\bf r}(\theta,\phi)=a_1({\rm sin}\theta{\rm cos}\phi)\hat{\i}+
a_2 ({\rm sin} \theta {\rm sin} \phi) \hat{\j}+
a_3 ({\rm cos} \theta) \hat{k}.
\end{equation}
\end{center}

We now define 
\[ E={\bf r}_{\theta} \cdot {\bf r}_{\theta}, \]
\[ F={\bf r}_{\theta} \cdot {\bf r}_{\phi},  \]
\[ G={\bf r}_{\phi} \cdot {\bf r}_{\phi}, \]
\[ L={\bf r}_{\theta \theta} \cdot {\bf n},  \]
\[ M={\bf r}_{\theta \phi} \cdot {\bf n}, \] 
\[ N={\bf r_{\phi \phi}} \cdot {\bf n},  \]
where 
\[ r_{\phi}={\partial {\bf r}}/{\partial \phi}, \]
\[ r_{\theta}={\partial {\bf r}}/ {\partial \theta}, \]
\[ r_{\phi \phi}={\partial^{2} {\bf r}}/{\partial \phi^{2}},  \]
\[ r_{\theta \theta}={\partial^{2} {\bf r}}/{\partial \theta^{2}}, \] 
\[ r_{\theta \phi}={\partial^{2} {\bf r}}/{{\partial \theta}{\partial \phi}}. \] 
The vector ${\bf n}$ represents the unit vector perpendicular to 
the surface and is defined as 
\[ \bf n = { {\bf r}_{\theta} \times {\bf r}_{\phi} } /
{\mid {\bf r}_{\theta} \times {\bf r}_{\phi} \mid}. \]

The four geometrical quantities can then be written as
\begin{equation}
S=\int \int \sqrt{EG-F^{2}} {\rm d}\theta {\rm d}\phi,
\end{equation}
\begin{equation}
C=\int \int \frac{k_{1}+k_{2}}{2} {\rm d} S,
\end{equation}
\begin{equation}
{\cal G}=\frac{-1}{4\pi} \int \int k_{1}k_{2} {\rm d} S,
\end{equation}
\begin{equation}
V=\frac{4}{3}\pi a_1 a_2 a_3. \; \;
\end{equation}

The principal curvatures of the ellipsoid are  $k_1$ and $k_2$, and the 
product and sum of these quantities are
\begin{equation}
k_{1}+k_{2}=\frac{EN+GL-2FM}{EG-F^{2}},
\end{equation}
\begin{equation}
k_{1}k_{2}=\frac{LN-M^{2}}{EG-F^{2}} \;\;.
\end{equation}

Three parameters are introduced, $H_{1}$, $H_{2}$ and $H_{3}$, 
which have dimensions of length: $H_1$=$3V/S$, $H_2$=$S/C$ and  $H_3$=$C/4\pi$. 
Combinations of these parameters provide two important \textit{shapefinders}, 
$K_{1}$ (planarity) and $K_{2}$ (filamentarity), which can be expressed as 
\begin{equation}
K_{1}=\frac{ {H}_{2}-{H}_{1} }{ {H}_{2}+{H}_{1} } 
\end{equation}
and
\begin{equation}
K_{2}=\frac{ {H}_{3}-{H}_{2} } { {H}_{3}+{H}_{2} } \;\;. 
\end{equation}
The vector $\mathbf{K}=(K_1,K_2)$ has an amplitude and direction which determine the shape of a certain 3D surface. An ideal pancake-like object presents one dimension which is much smaller than the others, so $H_1<<H_2\simeq H_3$ and consequently $\mathbf{K} \simeq(1,0)$. For an ideal filament, $H_1>>H_2\simeq H_3$ and so $\mathbf{K} \simeq (0,1)$. Considering surfaces like ribbons, the shapefinders have three distinct dimensions, i.e., $H_1<<H_2<<H_3$ and $\mathbf{K} \simeq (\alpha,\alpha)$ with $\alpha<1$. It is worth mentioning that, for a sphere, $H_1=H_2=H_3$ and hence $\mathbf{K}=(0,0)$. 

This formalism can be used to classify objects with different shapes, so we consider two morphologies:
\begin{itemize}
 \item objects with $K_1/K_2 >1$ are classified as pancakes.
 \item objects with $0 \le K_1/K_2 \le 1$ are classified as filaments.
\end{itemize}
The  range of $K_1/K_2$ for ribbons is somewhat arbitrary ($K_1/K_2\simeq 1$), so we decided to exclude this morphology from our classification.  
The shape statistics $K_1/K_2$, through the so-called ``shape spectrum'', 
was first applied to astronomy by B01.

\begin{figure}
\includegraphics[scale=0.5]{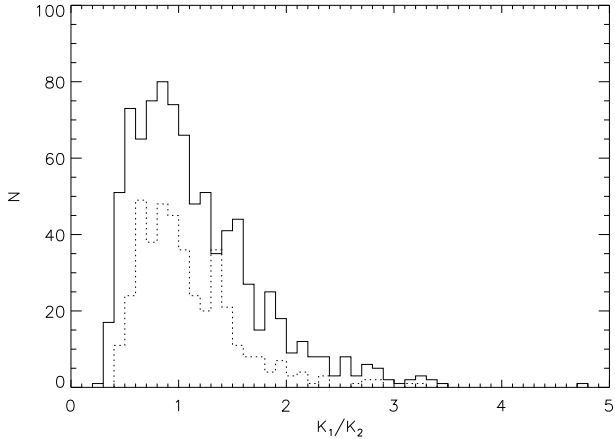} 
\caption{The histogram shows the distribution of $K_1/K_2$ for SDSS superclusters considering D1 (solid line) and D2 (dotted line). Both distributions present a median value around unity.}
\label{hist-k1k2}
\end{figure}

Figure~\ref{hist-k1k2} shows the distribution of the morphological parameter $K_1/K_2$. Table 1 presents some statistical properties of the supercluster morphologies considering the two threshold densities discussed in Section 3.1.

\begin{center}
\begin{table*}
 \begin{minipage}{150mm}
  \caption{Statistics for observed and simulated superclusters for the two threshold densities. The table shows the number of superclusters classified as filaments ($N_{f}$), pancakes ($N_{p}$), the mean number density ($\bar{n}_{SC}$) of superclusters and median values for total luminosity, richness, and $K_1/K_2$. For simulated superclusters, we also present our results in position space.}
  \begin{tabular}{ccccccccc}
 \hline
   Threshold & Sample & $N_{f}$ & $N_{p}$ & $\bar{n}_{SC}$(10$^{-5}$~h$^3$~Mpc$^{-3})$ & $\log(L_{tot}/L_{\odot})$ & $\log(R)$ & $K_1/K_2$ \\    
 \hline
D1  & SDSS-DR7 &  436 & 444 & 1.55 & 11.82 & 1.39 & 1.00  \\
(velocity space) &012.000  & 74 & 83 & 1.02 & 12.01 & 1.58 & 1.02 \\
 &012.100  & 70 & 75 & 0.94 & 12.02 & 1.60 & 1.01  \\
 &120.000  & 86 & 86 & 1.11 & 11.99 & 1.55 & 1.00  \\
 &201.000  & 76 & 81 & 1.02 & 12.01 & 1.60 & 1.04  \\ \hline
D2  & SDSS-DR7  & 204 & 212 & 0.74 & 12.07 & 1.64 & 1.01  \\
(velocity space) &012.000 & 31 & 27 & 0.37 & 12.29 & 1.85 & 0.92   \\
 &012.100 & 29 & 29 & 0.37 & 12.37 & 1.90 & 1.00 \\
 &120.000 & 45 & 23 & 0.44 & 12.28 & 1.83 & 0.88  \\
 &201.000 & 29 & 28 & 0.37 & 12.30 & 1.86 & 0.99 \\ \hline
 D1 & 012.000 & 98 & 90 & 1.22 & 11.97 & 1.55 & 0.97 \\
 (position space) &012.100 & 85 & 85 & 1.10 & 12.03 & 1.60 & 1.01  \\
 &120.000 & 82 & 109 & 1.24 & 12.00 & 1.57 & 1.08 \\
 &201.000 & 84 & 89 & 1.06 & 11.96 & 1.55 & 0.99 \\ \hline
D2 & 012.000 & 31 & 27 & 0.38 & 12.24 & 1.78 & 1.07  \\
(position space) &012.100 & 29 & 29 & 0.37 & 12.33 & 1.89 & 1.03 \\
 &120.000 & 34 & 27 & 0.40 & 12.24 & 1.78 & 0.97 \\
 &201.000 & 29 & 32 & 0.39 & 12.21 & 1.77 & 1.00 \\ \hline
\end{tabular}
\end{minipage}
\label{table1}
\end{table*}
\end{center}
\section{Results and Discussion}
\label{resultsanddiscussion}
 In this section, we investigate how overall properties of superclusters, like
their richness and total luminosity, correlate with their morphological
properties. We also repeat the analysis for structures extracted from
numerical simulations, with the main objective of comparing the supercluster
properties in the position and velocity spaces.

\subsection{SDSS superclusters}
\label{sdsssupeclusters}
Figure~\ref{hist-k1k2} and Table 1 indicate that we have essentially the same
number of objects classified as filaments or pancakes in our sample, for the two
density thresholds discussed here. We found 436 filaments and 444 pancakes for the threshold D1 and 204 filaments and 212 pancakes for the threshold D2. Hence, our result, does not confirm previous
works (e.g. B01), where a prevalence of objects classified as filaments was found. 

In order to investigate relations between morphology and properties of superclusters, the Spearman rank-order correlation coefficient, $r_s$,  was used to measure possible correlations \citep{Pressteal2007}. We have also computed the two-sided significance level of the null hypothesis of absence of correlation (or anti-correlation), $P(H_0)$; a small value of $P(H_0)$ is indicative of strong correlation (or anti-correlation). 

\begin{figure}
\begin{center}
\includegraphics[scale=0.5]{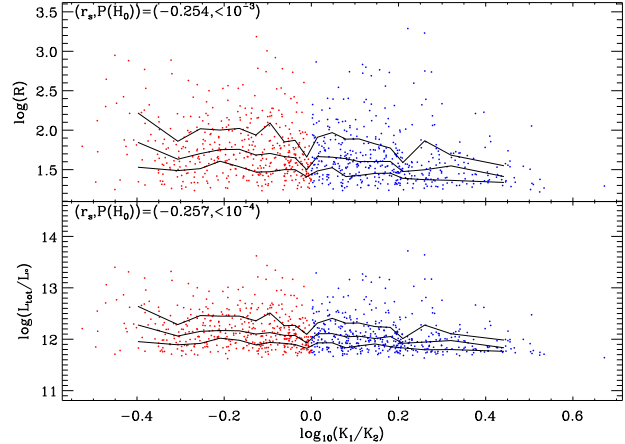} 
\caption{Richness (top) and total luminosity (bottom) of SDSS superclusters as a function of the morphological parameter $K_1/K_2$ for the density threshold D1.
The median and quartiles of the distribution are shown. } 
\label{morphology-k1k2-sdss}
\end{center}
\end{figure}

Figure~\ref{morphology-k1k2-sdss} shows the richness and total luminosity of SDSS superclusters as a function of the morphological parameter $K_1/K_2$, considering the threshold D1. The lines represent the median and quartiles for each bin of $K_1/K_2$. Despite the large scatter, there is a significant trend between  richness (and consequently total luminosity) and the morphological parameter $K_1/K_2$. In both cases we found $r_s \simeq -0.25$ and $P(H_0)<10^{-4}$, showing that filamentary structures tend to be richer and more luminous than pancakes. A similar behavior is found for the threshold D2, with a correlation coefficient $r_s \simeq -0.25$ and $P(H_0)<10^{-4}$ for both richness and total luminosity. These results indicate that the trends of richness and total luminosity with the morphological parameter are not strong but are statistically significant.

\begin{figure}
\begin{center}
\includegraphics[scale=0.5]{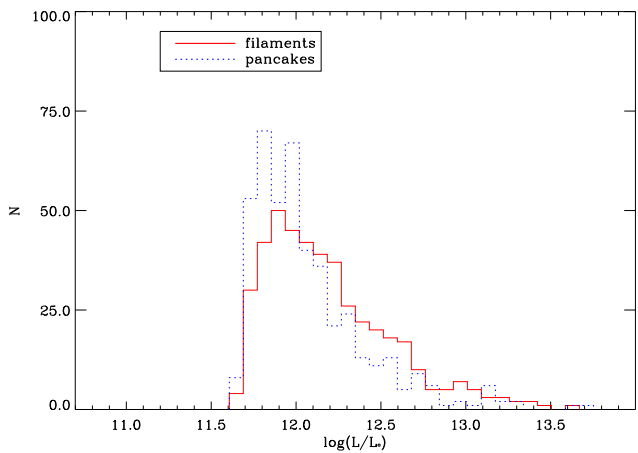} 
\caption{Observed luminosity distribution of filaments (continuous red histogram) and pancakes (dotted blue histogram), for density threshold D1.} 
\label{distra1}
\end{center}
\end{figure}

We have also compared the luminosity distributions of filaments
and pancakes through the Kolmogorov-Smirnov (K-S) test (Figure~\ref{distra1}). We conclude that their distributions are statistically distinct, presenting a K-S probability lower than $10^{-3}$ that the cumulative luminosity distributions of filaments and pancakes are drawn from the same distribution. A similar result is achieved using the threshold D2, in this case with a probability $<10^{-3}$. Indeed, Figure~\ref{distra1} suggests that the luminosity distribution of filaments is significantly broader than that of pancakes, resulting in a higher number of filaments at high luminosities, in agreement with Figure~\ref{morphology-k1k2-sdss}. A similar trend for superclusters classified as filaments to be richer and consequently more luminous had already been reported by E07b.

Considering now only the brightest structures, those with 
$\log(L/L_{\odot})>12.5$, we may notice the prevalence of filaments over pancakes: 60.4\% and 39.6\%, respectively, for the threshold D1 and  62.5\% and 37.5\%  for the threshold D2.

\begin{figure}
\begin{center}
\includegraphics[scale=0.6]{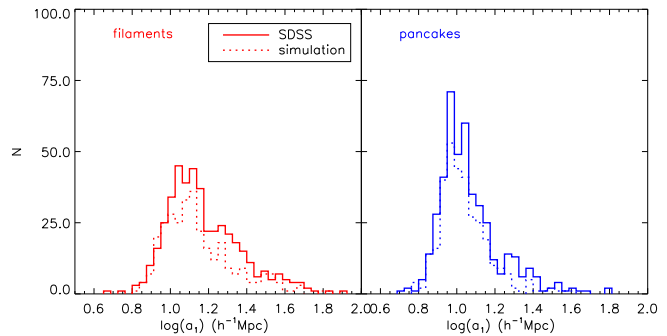} 
\caption{Distribution of the semi-major axis resulting from the ellipsoidal
fitting to the structures. Left: filaments, right: pancakes. Dotted lines  
correspond to the distributions obtained from the superclusters extracted
from the numerical simulations described in Section~\ref{simulations}. Results for the
density threshold D1.} 
\label{hist-a1-morph}
\end{center}
\end{figure}

Figure~\ref{hist-a1-morph} shows the distribution of the semi-major axis 
resulting from the ellipsoidal fitting to the structures. Filaments and pancakes
have different sizes, as demonstrated also by Figure~\ref{a1-k1k2}, with
filaments comprising most of the largest superclusters.

\subsection{Analysis of simulated superclusters}
\label{simulations}

We have repeated the above analysis for the four mock SDSS catalogues described in Section 2, applying the same procedures described in Sections~\ref{densityfield} and \ref{morphclass}. 

Table 1 shows the number of superclusters identified in the simulations, classified as filaments or pancakes, as well as the mean number density and median values of richness, total luminosity and $K_1/K_2$ for each sample of superclusters and threshold densities. Comparing the median values of total luminosity and richness from both threshold densities, slightly higher values are found in D2 because only richer and more luminous superclusters are identified using this threshold, increasing the median values.

\begin{figure}
\begin{center}
\includegraphics[scale=0.5]{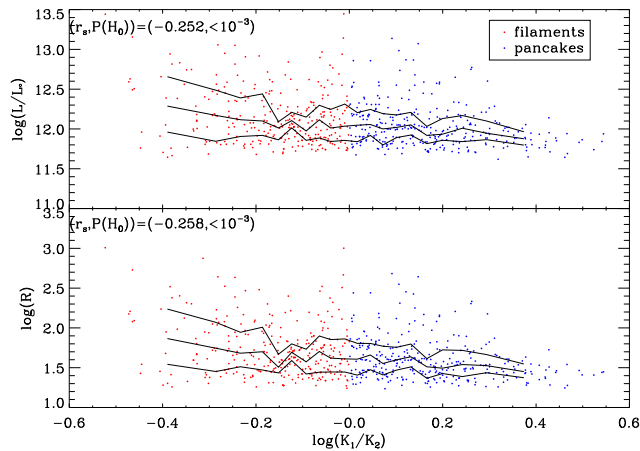} 
\caption{Richness (top) and total luminosity (bottom) of superclusters extracted from numerical simulations as a function of the morphological parameter $K_1/K_2$ for the density threshold D1. The median and quartiles of the distribution are shown. } 
\label{morphology-k1k2-LC}
\end{center}
\end{figure}

The same behaviour of observed richness and total luminosity with the morphological parameter $K_1/K_2$ is present in the simulations. Figure~\ref{morphology-k1k2-LC} shows the relation between richness and total luminosity and $K_1/K_2$ for the superclusters extracted from the simulations with density threshold D1. The value of the Spearman coefficient is $r_s=-0.25$ and $P(H_0)<10^{-4}$ considering the threshold D1, and $r_s=-0.25$ and $P(H_0)<10^{-4}$ for D2. This anti-correlation between richness and total luminosity and $K_1/K_2$ in the mock catalogues is very similar to that obtained for observed superclusters (Figure \ref{morphology-k1k2-sdss}), indicating a good agreement between simulations and observations. 

\begin{figure}
\begin{center}
\includegraphics[scale=0.5]{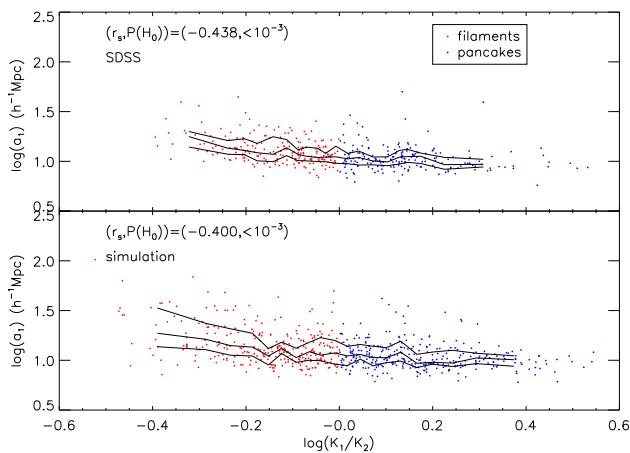} 
\caption{Semi-major axis resulting from the ellipsoidal
fitting to the structures as a function of the morphological parameter $K_1/K_2$ for the density threshold D1. The median and quartiles of the 
distribution are shown. Top: SDSS sample. Bottom: superclusters extracted
from the numerical simulations described in Section~\ref{simulations}.} 
\label{a1-k1k2}
\end{center}
\end{figure}

Peculiar velocities may have an important influence on supercluster identification. The observed redshift of extragalactic objects has a component produced by the Hubble flow plus a peculiar velocity component due to gravitational interactions. Consequently, there is a difference between measured distances in velocity and position spaces which produces some features in the velocity space such as the fingers-of-god and the Kaiser effect \citep[e.g.,][]{Bahcalletal1986,kaiser1987}. In order to study the influence of peculiar velocities on supercluster identification and morphology, we identified superclusters in the mock samples of \cite{Crotonetal2006} using two sets of redshifts available in the simulations: the first one including peculiar velocities (velocity space) and the second without them (position space). The influence of peculiar velocities on the identification and morphological classification of superclusters can be studied by comparing the properties of simulated superclusters in both spaces. Table 1 also presents the main features of simulated supercluster samples identified in velocity and position spaces. Qualitatively, results are similar for density thresholds D1 and D2.

\begin{figure}
\begin{center}
\includegraphics[scale=0.75]{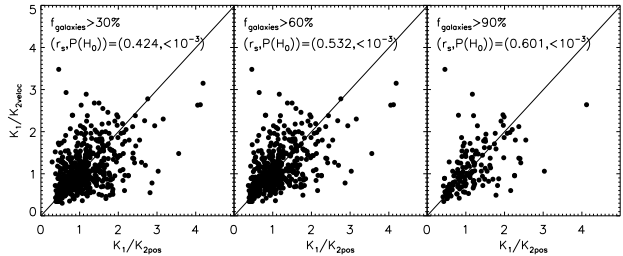} 
\caption{Comparison between the values of the morphological parameter of simulated superclusters identified in the velocity and position spaces. Each panel corresponds to different limits in the fraction of galaxies in common ($f_{g}$) in structures identified in the velocity and position spaces.} 
\label{comparison}
\end{center}
\end{figure}

\begin{figure}
\begin{center}
\includegraphics[scale=0.5]{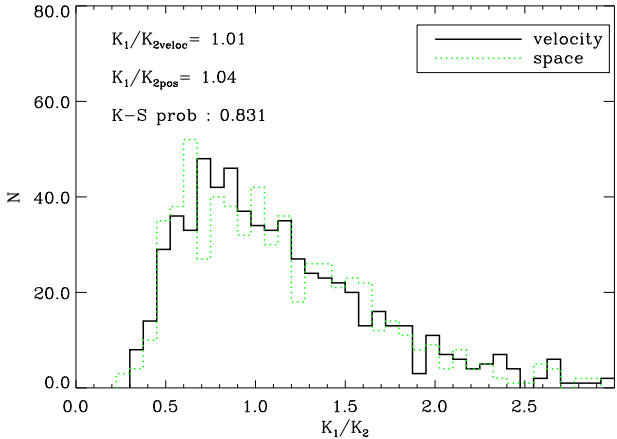} 
\caption{Distribution of the morphological parameter of superclusters identified in velocity (black) and position (green) spaces. The median values of $K_1/K_2$ (top left) are similar and the K-S test shows that the distributions are consistent with each other.}
\label{deltak1k2}
\end{center}
\end{figure}

Due to peculiar velocities, galaxies associated to a structure in position space may or may not be associated to the same structure identified in velocity space. In order to compare the morphologies of superclusters identified in both spaces, superclusters from position space were associated to velocity space ones according to the percentage of galaxies ($f_{g}$) in common (30\%, 60\% and 90\%). Figure~\ref{comparison} shows ${K_1/K_2}_{veloc}$ versus ${K_1/K_2}_{pos}$ for different percentages of galaxies in common. In all cases there is a significant correlation between the morphological parameter measured in both spaces, despite the large scatter, which increases as $f_{g}$ increases. Interestingly, there is no significant bias in the morphological parameter measured in velocity space compared with the values measured in position space. This is actually due to the high values of the smoothing parameter adopted here (see Section~\ref{densityfield}).
As another test of how morphology is distorted in redshift space, we have
considered galaxies in superclusters identified in redshift space and compared the morphology they trace in both velocity and position spaces. Figure~\ref{deltak1k2} shows the distribution of $K_1/K_2$ in velocity and position-spaces in this case. The median values of $K_1/K_2$ are similar (${K_1/K_2}_{veloc}=1.01$ and ${K_1/K_2}_{pos}=1.04$) in both spaces. A K-S test indicates that the two distributions are consistent with each other. Hence, this test also suggests that peculiar velocities are not important for our morphology measurements.

\begin{figure}
\begin{center}
\includegraphics[scale=0.57]{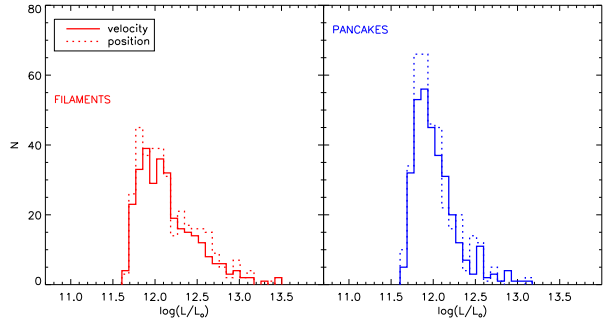} 
\caption{Luminosity distributions of filaments and pancakes identified in the mock catalogues, in velocity and position spaces, for the threshold D1.} 
\label{hist-morphology-D1}
\end{center}
\end{figure}

We now consider the luminosity distribution of superclusters identified in the velocity and position spaces in the mock catalogues. Figure~\ref{hist-morphology-D1} shows this distribution for filaments and pancakes for the density threshold D1. For each morphological class the distributions in both spaces are quite similar, as confirmed by the K-S test. It is worth mentioning that these distributions in velocity space resemble very much the observed distributions (Figure~9). This is confirmed by a K-S test comparison of the filament and pancake distributions with  K-S probability equals to 0.85 and 0.42, respectively, for the null
hypothesis that the data were drawn from the same distribution. Both probabilities show that the distributions are not statistically distinct.

The trend of the semi-major axis ($a_1$) of the ellipsoidal fitting of 
observed and simulated superclusters with the morphological parameter shown
in Figure~\ref{a1-k1k2} provides an indication that the largest structures 
tend to be filamentary. The trend is very similar for the observed and simulated
structures.

Our results indicate that simulated and observed superclusters present similar properties. This is also the case for the luminosity distributions of observed and simulated superclusters, as can be seen in Figures~\ref{distra1} and \ref{hist-morphology-D1}. However, E06 have found an absence of very luminous superclusters in simulations compared with observations. Indeed, a comparison between the observed and mock samples shows that objects brighter than $\log(L/L_{\odot})>12.5$ are at least twice more frequent in the observations than in the mock samples.

Additionally, by comparing the luminosity distribution of superclusters classified as filaments and pancakes, we have noticed that these classes have very different luminosity distributions. Since filaments tend to be richer and more luminous than pancakes, it is fair to suggest that these two morphological classes represent different evolutive dynamical stages of the large scale structure, with pancakes possibly evolving to filaments.

\subsection{Statistical reliability of supercluster identification}

Our method of supercluster identification is strongly dependent on the 
threshold density and on the smoothing parameter. Thus it is necessary to 
verify its robustness by checking  what fraction of our detections
is expected to be real or just statistical noise.

To investigate the expected number of random clumps, we follow the approach
proposed by \cite{Basilakos2003}. We ran a large number (200) of Monte Carlo
simulations where we randomized the equatorial coordinates of the 
galaxies, keeping their comoving distances and the 
sample boundaries in order to preserve the selection function. Since this
randomization destroys the SDSS clustering, the structures
identified in each simulation will be due to statistical noise. 
The probability of identifying real superclusters in our sample through
this procedure is then
\begin{equation}
P=1-\frac{N_{rand}}{N_{SDSS}},
\end{equation}
where $N_{rand}$ is the number of structures identified in the randomized 
samples and $N_{SDSS}$ is the number of structures identified in our original 
SDSS sample. A probability close to 1 means that the number of spurious objects
produced by our supercluster identification method is small. 

\begin{figure}
\begin{center}
\includegraphics[scale=0.5]{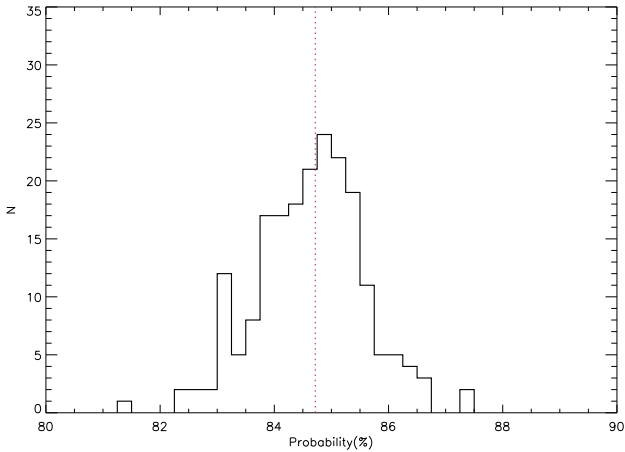} 
\caption{Distribution of the probability of identifying real superclusters for
the density threshold D1. The median value (red dotted line) is around 85\%.} 
\label{hist-random}
\end{center}
\end{figure}

\begin{figure}
\begin{center}
\includegraphics[scale=0.5]{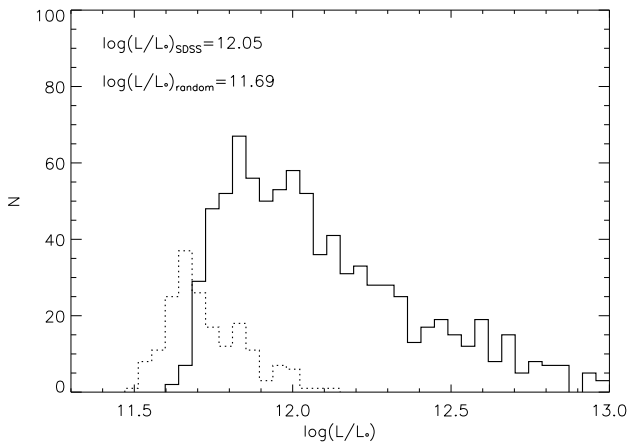} 
\caption{Luminosity distributions of SDSS superclusters (continuous line) and random superclusters (dotted line) for
the density threshold D1. The median of logarithmic values (top left) show that random superclusters present significantly lower luminosities than SDSS superclusters.} 
\label{hist-ltot-rand}
\end{center}
\end{figure}

This analysis was performed for the same threshold densities as 
in section \ref{kernel}. 
Figure \ref{hist-random} shows the resulting probability distribution obtained with the threshold D1, presenting a median probability $P$ around 85\%. For the threshold D2, the median probability is higher than 99.8\%. 
Figure~\ref{hist-ltot-rand} compares the luminosity distribution of our 
D1 SDSS superclusters with that obtained from the simulations. The result
clearly indicates that most random structures have luminosities significantly
lower than those of the SDSS superclusters. This result is also noticed by 
comparing the median values of the SDSS and random luminosity distributions.

The same morphological analysis as described in section \ref{morphclass} was 
applied to the structures identified in the Monte Carlo simulations.
No trends were found between the total luminosity or richness and the 
morphological parameter $K_1/K_2$, indicating that the results displayed in 
Figure \ref{morphology-k1k2-sdss} are indeed real and not produced by
random fluctuations or by the selection function of the sample.

Another test of robustness of the main trends identified in this work can
be done by considering only structures with total luminosities larger than
$10^{12}~L_\odot$, since our Monte Carlo simulations indicate that the
expected number of spurious structures in this case is only $\sim$3.7\%.
All our previous results are confirmed.

\section{Summary and conclusions}
\label{conclusion}

We have selected a volume-limited sample of galaxies from the SDSS in order to identify galaxy  superclusters, determine their morphologies and investigate some of their features. Our SDSS sample contains galaxies with $M_r<-21$ in the redshift range $0.04<z<0.155$ and covers stripes 10 to 37. We have also analyzed simulated light-cones based on a semi-analytic galaxy evolution model applied to the output of the Millenium Simulation \citep{Crotonetal2006}, to confront theory and observations and to examine the role of some systematic effects.

Superclusters were identified by the density field method with an Epanechnikov 
kernel, taking into account selection and boundary effects. By 
comparing the results obtained with this kernel with those obtained with a 
truncated Gaussian kernel, we show that the results are not strongly dependent on the 
kernel choice. The kernel smoothing parameter and the dimension of the density field cell were chosen as $\sigma$=8~h$^{-1}$Mpc and $l_{cel}$=4~h$^{-1}$Mpc. Two threshold densities were chosen to evaluate their influence on supercluster identification and morphology. The first maximizes the number of structures (D1) and the second is chosen by limiting the size of the largest superclusters to $\sim$120~h$^{-1}$Mpc (D2). We found that, at least qualitatively, our results do not depend on the density threshold used. Each supercluster is characterized by its richness and total luminosity, as well as by a morphological parameter determined through Minkowski Functionals, which allows their classification as filaments or pancakes. 

Following E07b, we have found significant correlations of the morphological parameter $K_1/K_2$ with richness and total luminosity in both the observed and mock supercluster catalogues, indicating that filaments tend to be richer and consequently more luminous than pancakes. 

To evaluate the influence of peculiar velocities on supercluster morphology, we have used mock catalogues to examine structures identified in the velocity and  position spaces. We conclude that peculiar velocities do not play a significant role in our results, probably due to the large kernel smoothing length adopted in this work. 

We have found a trend between supercluster total luminosity (or richness) and morphology, with filaments being in the mean more luminous than pancakes. A similar behaviour was found by analyzing mock catalogues. The analysis of Monte Carlo
simulations of randomized galaxy distributions indicates that these trends are
real and are not produced by random fluctuations or selection effects.

Finally, we compared the luminosity and size distributions of filaments and pancakes. We have found that they are significantly different, with filaments presenting a broader distribution of sizes and luminosities. Again, similar results were obtained by the analysis of the mock catalogues, showing that filaments and pancakes represent distinct morphological classes of superclusters in the Universe. Also, since filaments tend to be richer, more luminous and larger than pancakes, it is plausible to think that pancakes evolve towards filaments.

\section*{Acknowledgments}
We are grateful to the referee, Spyros Basilakos, for his constructive and
very helpful comments.
MVCD thanks CAPES for a scholarship that allowed him to develop this project. LS acknowledges FAPESP and CNPq for their support to his research. FD and LS acknowledge the support of the Brazilian-French collaboration CAPES/Cofecub (444/04). We also wish to thank the team of the Sloan Digital Sky Survey (SDSS) for their dedication to a project which has made the present work possible.

\label{lastpage}
\appendix
\section{The influence of the smoothing kernel on supercluster properties}
\label{apend_a}
The kernel in Equation 2 plays a fundamental role in the density field 
calculation and therefore it is important to verify how sensitive our 
results are to the kernel choice. We address this point by comparing  
supercluster properties obtained with the Epanechnikov kernel with their
properties obtained with a truncated Gaussian kernel, defined as
\begin{equation}
 K(r,\sigma)=\frac{1}{2\pi}e^{\frac{-r^2}{2\sigma^2}},
\label{kernel_gauss}
\end{equation}
with a cutoff at 3-$\sigma$.

\begin{figure}
\begin{center}
\includegraphics[scale=0.5]{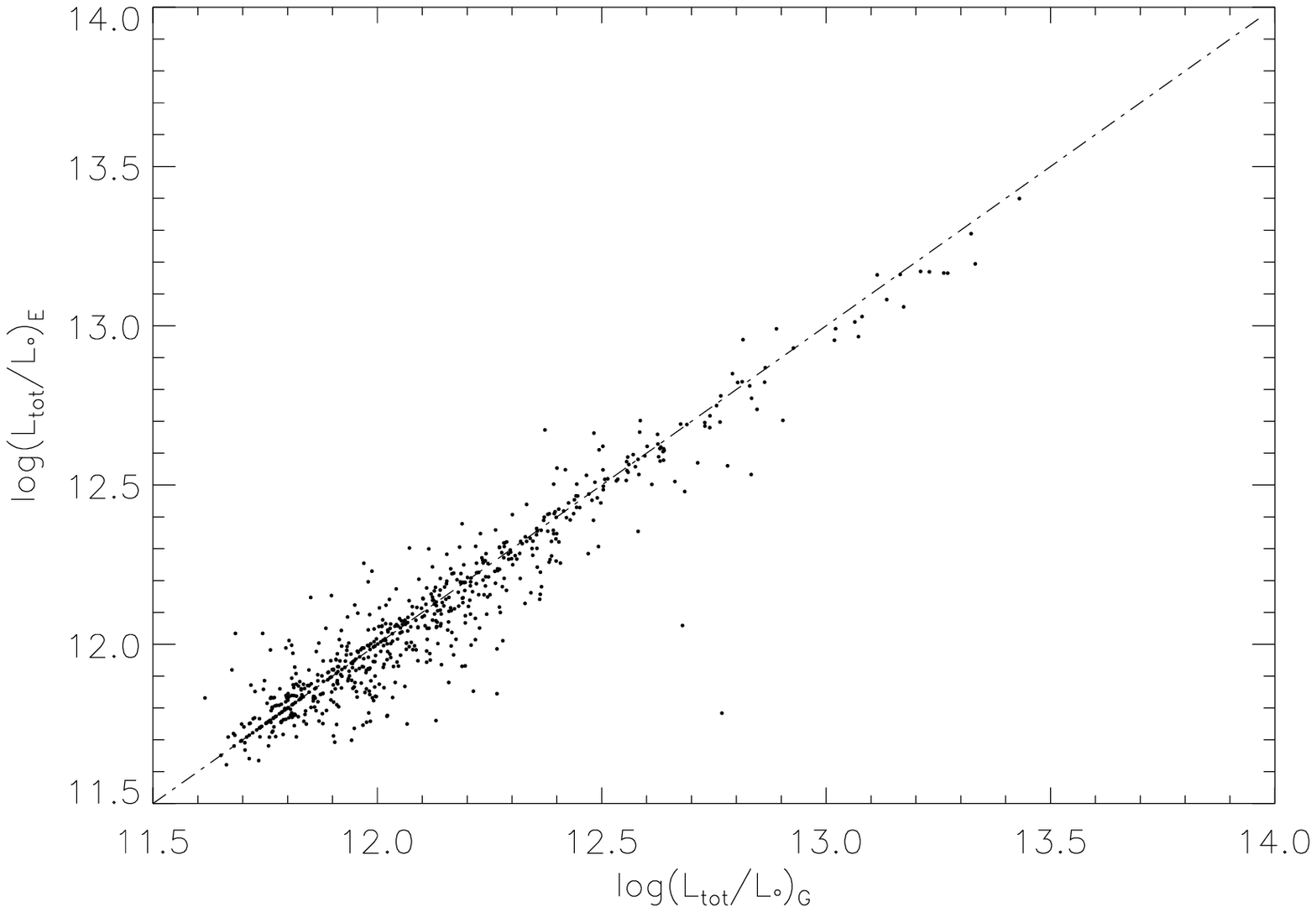} 
\caption{Comparion between the total luminosity of superclusters identified using the Gaussian ($log(L/L_{\odot})_{G}$) and Epanechnikov ($log(L/L_{\odot})_{E}$) kernels with parameters described in the text.} 
\label{ltot-1-2.4}
\end{center}
\end{figure}

For this exercise the density field was calculated as described in Section 
\ref{kernel} but with a smoothing parameter $\sigma$=2.4 h$^{-1}$Mpc for the 
truncated Gaussian model. This value leads to 1038 superclusters for the 
threshold D1 with luminosities similar to those discussed in Section 
\ref{sdsssupeclusters}, as shown in Figure  \ref{ltot-1-2.4}. A similar 
result is obtained by comparing the richness of superclusters obtained
with the two kernels.

In order to compare the structures obtained with these kernels, we have matched the two supercluster catalogues by assuming that the distance between the object center-of-mass in both catalogues is lower than 8 h$^{-1}$Mpc. In this way, we have identified 368 objects in common.

\begin{figure}
\begin{center}
\includegraphics[scale=0.5]{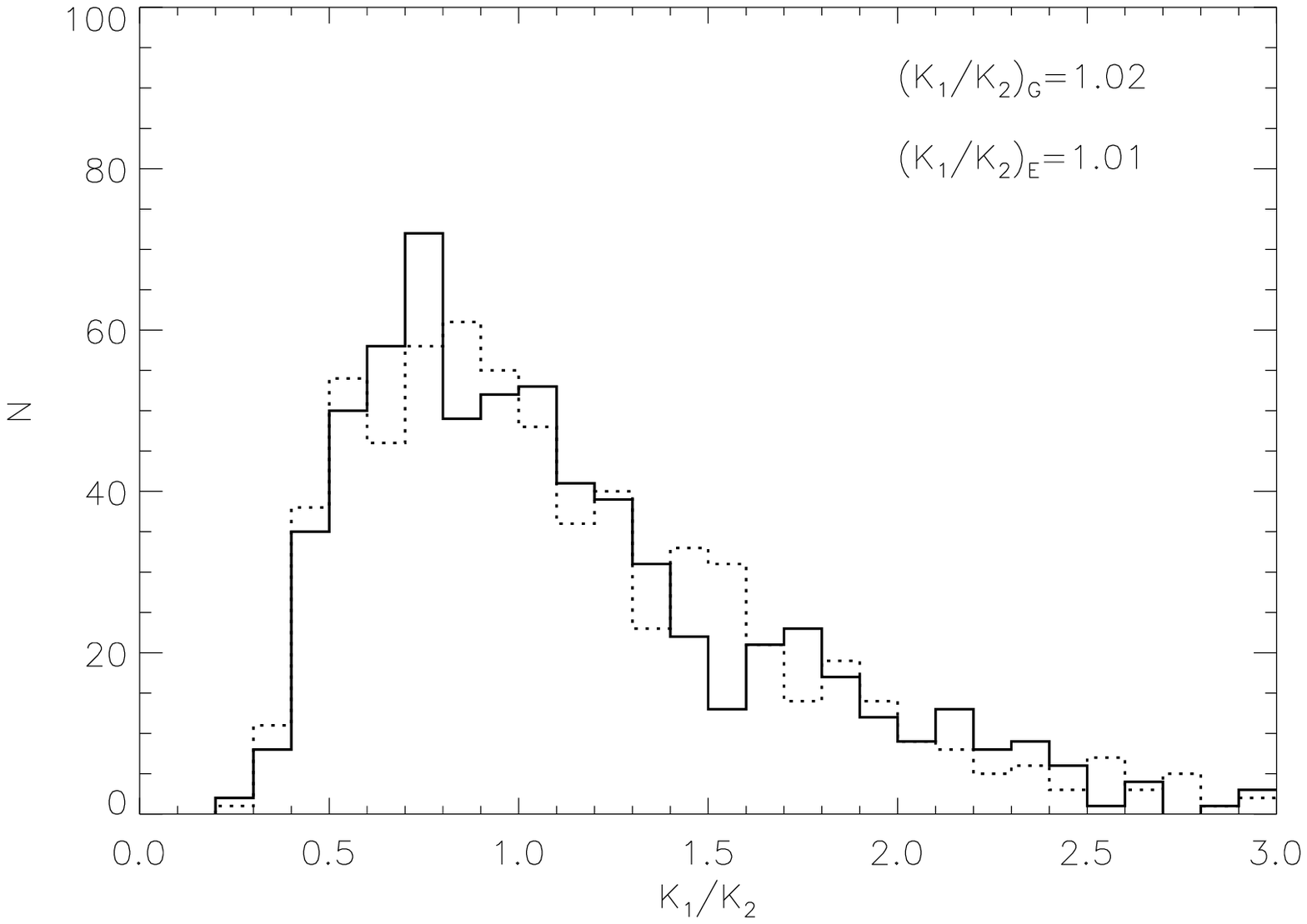} 
\caption{Distribution of the morphological parameter $K_1/K_2$ for 
superclusters identified using the Epanechnikov (dotted line) and Gaussian (continuous line) kernels. At the top right we show the median values of each 
distribution.} 
\label{hist-k1k2-kernels}
\end{center}
\end{figure}

In Figure \ref{hist-k1k2-kernels} we compare the distributions of the 
morphological parameter $K_1/K_2$ obtained with the two kernels. The 
distributions have similar medians and the K-S test did not distinguish them. The same trends present in Figure \ref{morphology-k1k2-sdss} were found using the Gaussian kernel, with a correlation coefficient $r_s\simeq -0.18$ and $P(H_0)<10^{-3}$ for both luminosity and richness.
Similar results are obtained with the threshold D2. We conclude that 
the trends described in the text are
actually robust with respect to the kernel choice, as far as sensible 
kernels are used.

\end{document}